\documentclass[final,3p,times,authoryear]{elsarticle}

\usepackage{natbib}
\usepackage{graphicx}
\usepackage{amsmath,amsfonts, amsthm, mathrsfs, amssymb, bbm}
\usepackage{times}

\begin{document}

\begin{frontmatter}

\title{Shepherd Model for Knot-Limited Polymer Ejection from a Capsid} 
\author[PED]{Tibor Antal}
\ead{tibor\_antal@harvard.edu}
\author[BU]{P. L. Krapivsky}
\ead{paulk@bu.edu}
\author[BU]{S. Redner}
\ead{redner@bu.edu}
\address[PED]{Program for Evolutionary Dynamics, Harvard University,
  Cambridge, MA~ 02138, USA}
\address[BU]{Center for Polymer Studies and Department of Physics, Boston University, Boston, MA~ 02215, USA}


\begin{abstract}
  We construct a tractable model to describe the rate at which a knotted
  polymer is ejected from a spherical capsid via a small pore.  Knots are too
  large to fit through the pore and must reptate to the end of the polymer
  for ejection to occur.  The reptation of knots is described by symmetric
  exclusion on the line, with the internal capsid pressure represented by an
  additional biased particle that drives knots to the end of the chain.  We
  compute the exact ejection speed for a finite number of knots $L$ and find
  that it scales as $1/L$. We establish a mapping to the solvable zero-range
  process. We also construct a continuum theory for many knots that matches
  the exact discrete theory for large $L$.
\end{abstract}

\begin{keyword}
Virus ejection \sep 
Exclusion process \sep
Stochastic effects
\end{keyword}

\end{frontmatter}

\section{Introduction}

The two basic steps in bacterial infection are the initial injection of the
viral DNA into a bacteriophage capsid \citep{RB78,KTBG01,PKP03} and then the
ejection of the DNA into a host cell \citep{IGP06,CE}.  The processes underlying
these two steps are both phenomenologically rich and incompletely understood.
In packaging DNA into a capsid, both the strong repulsion from the bending of
the DNA and the stress caused by the capsid being smaller than the
persistence length have to be overcome.  The pressure associated with this
capsid packaging can be in the range of tens of atmospheres \citep{S01,GK09}.
These forces that repel the polymer during packaging are overcome by a
specifically-designed motor protein \citep{IGP06}.

The complementary ejection of the DNA into the host cell through small pores
in the capsid is driven by the osmotic pressure exerted by the capsid and the
internal stresses that have been built up in the highly-confined DNA polymer
chain \citep{IGP06,GK09}.  Because of its central importance in the life
cycle of viruses, DNA ejection from a capsid has been intensively studied
\citep{IGP06}.  Most theoretical treatments have investigated the role of
microscopic mechanisms that arise naturally from the combined effects of
osmotic pressure and bending energy.

Recent experiments and simulations suggest that viral DNA may become knotted
({\it i.e.}, a closed chain with a knot cannot be smoothly deformed into a
circle) as it is being tightly packed in the capsid \citep{A02,A05}.  These
knots are apparently large enough to prevent the DNA from being completely
ejected being from the capsid \citep{MHRL05}.  If this steric hindrance is
fully operative, then knots would have to unravel for complete ejection.
Because the driving forces are pushing the chain out of the capsid, the only
way for complete ejection to occur is for the knot to reptate along the chain
and ultimately unravel when the end of the chain is reached.  This mechanism
and its role on polymer ejection has recently been explored by numerical
simulation of a bead-spring model of a flexible polymer together with a
coupling to a background solvent \citep{matthews09}.  This study provides
detailed, but nevertheless still qualitative results, for the time dependence
of the fraction of the chain that remains within the capsid, as well as the
dependence of the ejection time on the number and type of knots on the
polymer.

In this work, we propose a coarse-grained model to capture the role of knots
on the ejection process (Fig.~\ref{fig:illust}).  In this model, the
reptation of knots is described by the symmetric exclusion process
\citep{SEP,Ligg85,schutz00}, while the point at the interface between in
the interior and exterior of the capsid undergoes a biased motion and also
satisfies the exclusion constraint.  We may view this biased particle as a
``shepherd'' that pushes the knots (``sheep'') to the end of the chain where
they can unravel.  Once all the knots have unraveled, the polymer can be
completely ejected.

In the next section, we present the details of this shepherding model.  In
Sec.~\ref{discrete}, we treat the ejection of a polymer chain with a small
number of knots.  For this discrete system, we write and solve the master
equations that describe the position of the knots along the chain.  This
solution gives the exact result that the velocity of a strongly-biased
shepherd equals $1/L$, where $L$ is the number of knots.  An unexpected and
simplifying feature of the strongly-biased limit is that the velocity and
diffusion coefficient of the shepherd do not depend on microscopic rates.  We
also establish a mapping to the solvable zero range process \citep{evans05},
and we use this mapping to obtain further exact results for the diffusion of
the shepherd.  Finally, we investigate a semi-infinite system with a finite
density of knots (Sec.~\ref{cont}).  Here we apply a continuum approach to
solve for the knot density profile, from which the displacement of the
shepherd grows as $\sqrt{A\, t}$, with a calculable amplitude $A$.

\section{Model}
\label{model}

To present our model, it is convenient to employ a reference frame that is
fixed along the length of the polymer.  The chain can be divided into a
portion that has been ejected from the capsid and the portion that remains in
the interior.  The basic feature of our modeling is that ejection may be
hindered by the existence of knots along the portion of the polymer that is
still inside the capsid.  We assume that these knots are sufficiently large
that they cannot pass through the pore in the capsid \citep{matthews09}.
Because of this restriction, some other relaxation mechanism is needed to
allow for complete ejection of a polymer.

\begin{figure}[ht]
\centering
\includegraphics[width=0.45\textwidth]{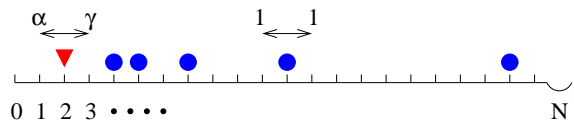}
\caption{Schematic illustration of the shepherding process.  The motion of
  the shepherd ($\blacktriangledown$) is biased to the right, while the knots
  ($\bullet$) are represented as localized particles that hop symmetrically
  on the integer interval $[0,N]$.  Upon reaching the end of the chain, the
  knot has unraveled; this event is represented as an absorbing boundary at
  $N$.  }
\label{fig:illust}
\end{figure}

One such mechanism is provided by the possibility that knots are not static
but move along the polymer chain by reptation \citep{matthews09}.  When a
knot reptates to the free end of the polymer, it disappears because the
entanglement of the knot is released.  If the chain contains multiple knots,
then we assume that they cannot pass through each other; instead they must
individually reptate to the end of the chain and unravel in sequence.

Our main focus is the position of the point that separates the ejected
portion of the polymer from the portion inside.  We model the position of
this dividing point by an effective particle (a ``shepherd'') that undergoes
biased diffusion along the chain, with rates $\alpha$ and $\gamma>\alpha$ to
the left and right, respectively.  The bias of the shepherd corresponds to
the polymer being pushed out of the capsid by osmotic pressure and the
release of bending energy.  The shepherd therefore rectifies the diffusive
motion of the knots so that the polymer will be ejected.  When the shepherd
reaches the right end of the polymer, the ejection is complete.  In the
spirit of minimalism, we assume that this bias is constant throughout the
ejection process; in reality the bias will decrease with time as the ejection
of the polymer relaxes the stresses that drives the ejection.

We assume that the chain initially contains a specified number of knots that
are randomly interspersed along the chain. The ejected portion is necessarily
free of knots, while the complementary portion of the chain remains knotted
(except possibly near the very end of the ejection process).  In the spirit
of a coarse-grained model, we ignore the topology of the knots and replace
them by localized defects.  Each defect can move equally likely in both
direction along the chain due to thermal noise, and hence the reptation of
the knot is modeled as a symmetric random walk.  Different type of knots
could be described by different diffusion coefficients, however, we restrict
our discussion to a single type of knot, where all knots have the same
hopping rates.  In addition to the random motion of the knots, there is also
an exclusion constraint because knots cannot pass through each other or pass
by the shepherd (corresponding to the disallowed event of a knot passing
through the capsid pore).  The knots can be viewed as a flock of diffusing
and mutually repelling excitations that are pushed to the end of the interval
by an advancing shepherd.  When each knot reaches the right end of the chain
it simply disappears, corresponding to an absorbing boundary condition.  When
the shepherd reaches the end of the chain (Fig.~\ref{fig:illust}), the
polymer has been completely ejected.

Because of its bias, the shepherd rectifies the diffusive motion of the
knots.  This rectification of thermal noise underlies many microscopic
biological processes, such as the motion of motor proteins \citep{motors},
and the chaperon assisted translocation of polymers across a membrane
\citep{OCA07}.  In our model of polymer ejection, the rectification is
mediated by the exclusion constraint of the knots, leading to an exclusion
process in which a single particle is subject to a bias that acts indirectly
on all other particles.  A two-sided version of this model with a biased
particle caged in between unbiased particles has been studied previously
\citep{BOMM,BOMR,L98}; this model mimics, {\it e.g.}, the field-driven motion
of a charged particle that is immersed in a lattice gas of neutral particles.
It is worth mentioning that various embellishments of idealized exclusion
processes have helped to understand a variety of biological processes and
have raised new theoretical questions about the exclusion process itself
\citep{RFE,SK,DSZ,spiders}.

\section{Discrete Formulation}
\label{discrete}

We first study the case of a small number of knots by writing the discrete
master equations that describe the probability distribution for their
positions.  These master equations turn out to be soluble, from which we can
extract the speed of the shepherd in the steady state.

\subsection{Single Knot}

\begin{figure}[ht]
\centering
\includegraphics[width=0.45\textwidth]{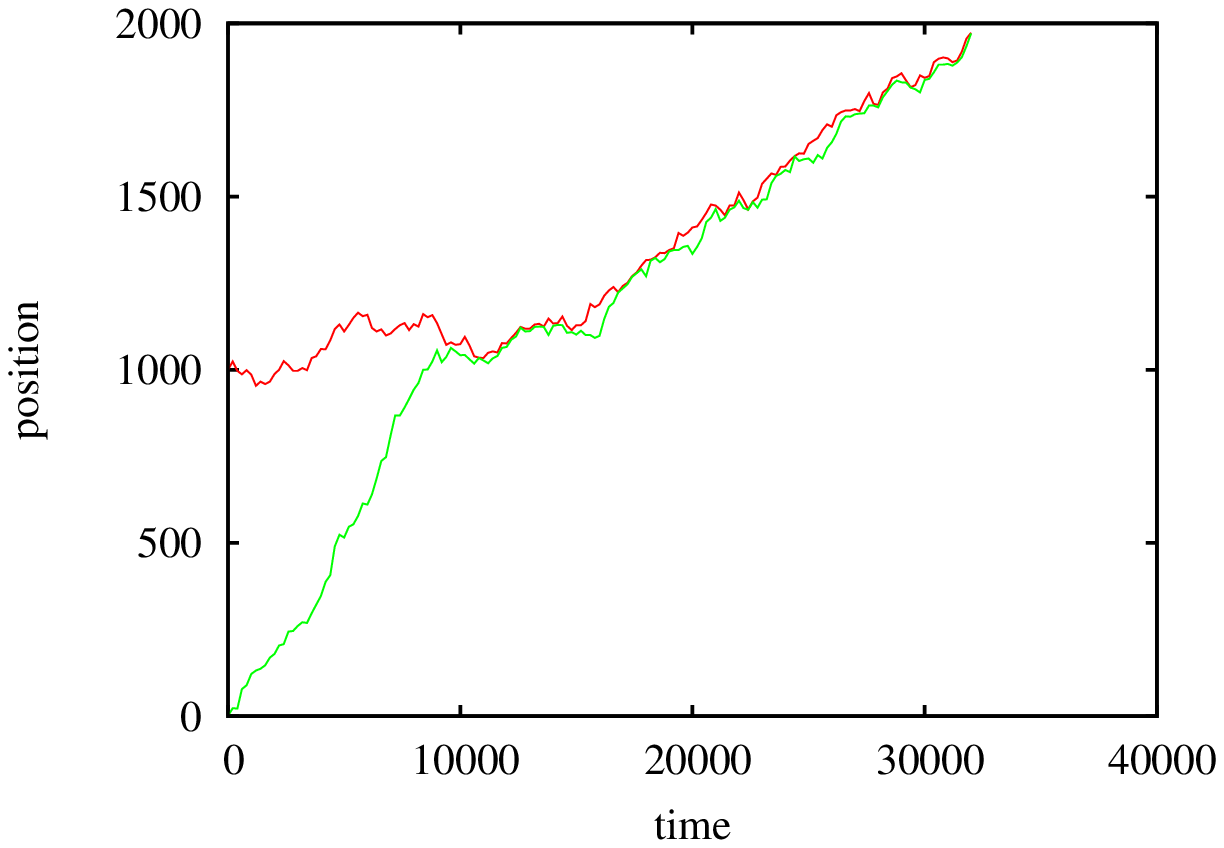}
\includegraphics[width=0.45\textwidth]{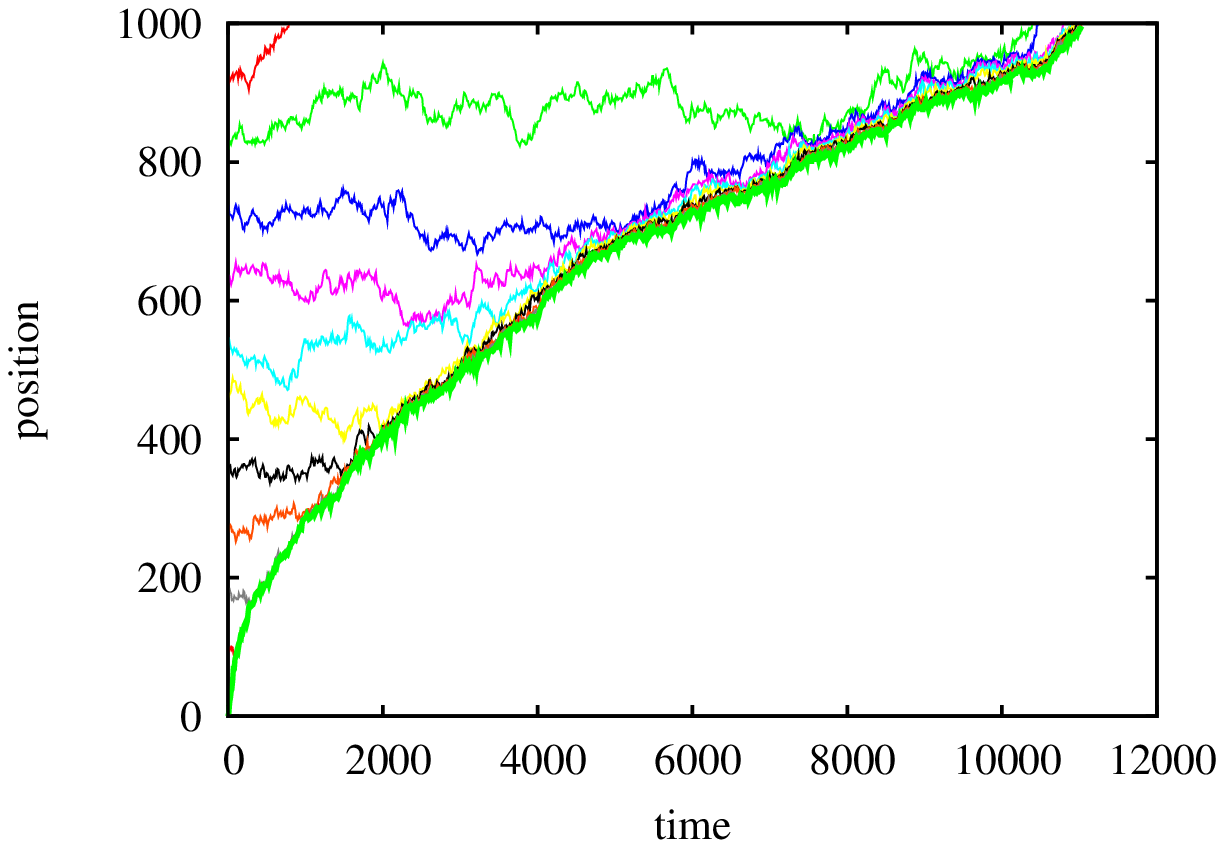}
\caption{(Left) Time dependence of the positions of the shepherd (lower) and
  a single knot (upper), when the shepherd hopping rates to the right and
  left are $\gamma=1.1$ and $\alpha=1$, respectively. (Right) Time dependence
  of the positions of the shepherd (lower) and 10 knots (upper) on a chain of
  length 1000, when the shepherd hopping rates to the right and left are
  $\gamma=2$ and $\alpha=1$, respectively. }
\label{xt}
\end{figure}

To determine the ejection speed of a polymer that contains a single knot,
note that before the shepherd first reaches the knot, its speed is simply
$v_0=\gamma-\alpha$ (Fig.~\ref{xt}, left).  Subsequently, the knot and the
shepherd stay close to each other because of the bias of the shepherd toward
the knot.  As a result, the speed of their mutual advance is less than $v_0$.
To obtain this speed, we define $P(n)$ as the probability that the number of
empty sites between the shepherd and the first knot is $n$.  We call this
number of vacancies between the two particles as the ``gap''.  For gap size
$n\geq 0$, $n$ can increase to $n+1$ with rate $1+\alpha$, either by the knot
hopping one step to the right or the shepherd hopping one step to the left
with rate $\alpha$, respectively.  Similarly, for $n\geq 1$, the gap size
$n$ decreases to $n-1$ with rate $1+\gamma$.  When $n=0$, the gap size can
only increase with rate $1+\alpha$.

Because the shepherd is driven toward the knot for $\gamma>\alpha$, the
distance distribution between these two entities reaches a steady state.  In
this state, the two transition rates $(1+\alpha)P(n)$ and $(1+\gamma)P(n+1)$
are equal and give the recursion, for $n\geq 0$,
$(1+\alpha) P(n) = (1+\gamma) P(n+1)$
for $n\ge 0$, with solution
\begin{equation}
\label{Pn}
  P(n) = \frac{\gamma-\alpha}{\gamma+1} \left( \frac{1+\alpha}{1+\gamma} \right)^n~.
\end{equation}
Here the overall amplitude is determined by the normalization condition
$\sum_{n\geq 0} P(n)=1$.

Since the shepherd can always hop to the left (with rate $\alpha$), but it can
hop to the right only if its right neighbor is empty (probability $1-P_0$) ,
the speed of the shepherd for the case of one knot is
\begin{equation}
\label{oneknotspeed}
 V_1 = \gamma(1-P_0) - \alpha = \gamma\frac{1+\alpha}{1+\gamma} - \alpha\,,
\end{equation}
In the limit of an infinitely biased shepherd ($\gamma\to\infty$),
Eq.~\eqref{oneknotspeed} reduces to $V_1=1$.

\subsection{Several Knots} 

For $L$ knots their configuration may be specified by the set of separations
$\mathbf{n}\equiv (n_1, n_2,\ldots, n_{L})$, where $n_i$ is the number of
vacancies to the left of the $i^{\rm th}$ knot.  The stationary probability
$P(\mathbf{n})$ for a configuration specified by $\mathbf{n}$ satisfies the
equation
\begin{equation}
\label{bulk}
\begin{split}
  \frac{dP(\mathbf{n})}{dt} &= -(\alpha+\gamma+2L)P(\mathbf{n}) 
+ \gamma P(n_1+1)+\alpha P(n_1-1)\\
& + \sum_{k=1}^{L-1} \Big[ P(n_k-1, n_{k+1}+1)
 + P(n_k+1, n_{k+1}-1) \Big] +P(n_L-1) + P(n_L+1) =0\,.
\end{split}
\end{equation}
The scalar arguments of $P$ on the right-hand side indicate that only the
components of the separation vector that have been changed from those in
$\mathbf{n}$.  Equation~\eqref{bulk} is valid for configurations in which the
separations between consecutive knots are all at least one ($n_k\ge 1$ for
every $k$).  For configurations where some $n_k=0$ ({\it i.e.}, zero
separation between adjacent knots), the relevant separations cannot decrease.
In such cases and independently for each $k$, the following terms are missing
({\it i.e.}, they should be subtracted) from the right-hand side of
\eqref{bulk}
\begin{equation}
\label{zerosep}
 \begin{tabular}{ll}
 $-(1+\gamma)P(\mathbf{n}) + \alpha P(n_1-1) + P(n_1-1, n_2+1)$  ~&\text{for}~ $k=1$\\
 $-2P(\mathbf{n}) + P(n_{k-1}+1,n_k-1)+P(n_k-1,n_{k+1}+1)$ ~&\text{for}~ $2\le k\le L-1$\\
 $-2P(\mathbf{n}) + P(n_{L-1}+1,n_L-1) + P(n_L-1)$  ~&\text{for}~ $k=L$.\\
 \end{tabular}
\end{equation}
(Here we formally allow $n_k$ to take the value $-1$ for convenience.)~ By
solving the system of equations \eqref{bulk} and \eqref{zerosep} for a small
number of knots, a simple pattern emerges that suggests that the exact
solution factorizes and individual gaps are distributed according to
exponential distributions
\begin{equation}
\label{prodanz}
 P(\mathbf{n}) = \prod_{k=1}^L (1-z_k)\,z_k^{n_k}~.
\end{equation}
(The prefactor in \eqref{prodanz} assures that normalization is obeyed: $\sum
P(\mathbf{n})=1$.)  Substituting ansatz \eqref{prodanz} into the bulk and
boundary equations, Eqs.~\eqref{bulk} and \eqref{zerosep}, we obtain
\begin{equation}
\label{bulksim}
 \alpha+\gamma+4 = \gamma z_1 + \frac{\alpha+z_2}{z_1} + \frac{z_{L-1}+1}{z_L} + z_L
 +\sum_{k=2}^{L-1} \frac{z_{k-1}-2z_k+z_{k+1}}{z_k},
\end{equation}
as well as 
\begin{equation}
\label{recursion}
 \begin{tabular}{ll}
 $z_1(1+\gamma)=\alpha+z_2$ , ~&\text{for}~ $k=1$\\
 $2z_k=z_{k-1}+z_{k+1}$, ~&\text{for}~ $2\le k\le L-1$\\
 $2z_L=z_{L-1}+1$ , ~&\text{for}~ $k=L$\\
 \end{tabular}
\end{equation}
When the conditions \eqref{bulksim} and \eqref{recursion} are both satisfied
the solution is stationary.  Equation \eqref{recursion} represents a linear
recursion with the unique solution
\begin{equation}
\label{solz}
z_k = \frac{\alpha L+1+(\gamma-\alpha)(k-1)}{\gamma L+1}~.
\end{equation}
This form for $z_k$ also satisfies the bulk equation \eqref{bulksim}, which
can be verified by substitution.


The speed of the shepherd can be calculated in the same way as in the case of
a single knot.  The shepherd can always jump to the left, but it can jump to
right only if its right neighbor is empty, that is $V_L = \gamma
\mathrm{Prob}(n_1\ge 1) - \alpha$.  The probability that the first gap has
size of at least one is
\begin{equation}
  \mathrm{Prob}(n_1\ge 1) = (1-z_1) \sum_{n_1=1}^\infty  z_1^{n_1} = z_1~.
\end{equation}
Hence the speed of the shepherd is
\begin{equation}
\label{speed}
V_L = \frac{\gamma-\alpha}{\gamma L+1}~.
\end{equation}

Equation \eqref{prodanz} shows that the probability distribution for the gap
size $\ell_k=1+n_k$ between the $(k-1)^{\rm st}$ and $k^{\rm th}$ knots is
$(1-z_k)\,z_k^{\ell_k-1}$ and therefore the average distance is
\begin{equation}
  \langle \ell_k\rangle = \sum_{\ell_k\geq 1} \ell_k\,(1-z_k)\,z_k^{\ell_k-1}=\frac{1}{1-z_k} = \frac{\gamma L +1}{\gamma -\alpha}\,\frac{1}{L-k+1}
\end{equation}
Since gaps are independent, the average total size ({\it i.e.}, the average
distance between the shepherd and the $L^{\rm th}$ knot) is
\begin{equation}
\label{av-size}
\langle \ell\rangle = \sum_{k=1}^L\langle \ell_k\rangle = 
\frac{\gamma L +1}{\gamma -\alpha}\,H_L
\end{equation}
where $H_L=\sum_{1\leq j\leq L} j^{-1}$ are harmonic numbers. Using the
asymptotic $H_L\simeq \ln L$ when $L\gg 1$ we see that the average gap size
scales according to
\begin{equation}
\label{asymp-size}
\langle \ell\rangle\simeq 
\frac{\gamma}{\gamma -\alpha}\,L\ln L
\end{equation}
when the number of knots is large. Similarly, we use the independence of gaps
to compute the variance of the total size
\begin{equation}
\langle \ell^2\rangle - \langle \ell\rangle^2 = \sum_{k=1}^L
\left[\langle \ell_k^2\rangle - \langle \ell_k\rangle^2\right] = \sum_{k=1}^L
\frac{z_k}{(1-z_k)^2} 
\end{equation}
which simplifies [upon using  \eqref{solz}] to 
\begin{equation}
\label{var-size}
\langle \ell^2\rangle - \langle \ell\rangle^2 = 
\left(\frac{\gamma L +1}{\gamma -\alpha}\right)^2H_L^{(2)}-
\frac{\gamma L +1}{\gamma -\alpha}\,H_L
\end{equation}
where $H_L^{(2)}=\sum_{1\leq j\leq L} j^{-2}$. For large number of knots
$H_L^{(2)}\to \pi^2/6$ and therefore asymptotically 
\begin{equation}
\langle \ell^2\rangle - \langle \ell\rangle^2 = \frac{\pi^2}{6}
\left(\frac{\gamma}{\gamma -\alpha}\right)^2 L^2 + \mathcal{O}(L\ln L)
\end{equation}
Hence the relative fluctuations of the total size diminish, albeit slowly as
the inverse logarithm of the total number of knots:
\begin{equation}
\frac{\sqrt{\langle \ell^2\rangle - \langle \ell\rangle^2}}{\langle \ell\rangle} 
\sim  \frac{1}{\ln L}
\end{equation}

In the special case of an infinitely biased shepherd ($\gamma=\infty$), the
shepherd is always adjacent to the first knot, so that $n_1=0$ and the
stationary probability \eqref{prodanz}, \eqref{solz} becomes
\begin{equation}
 P(\mathbf{n}) = \prod_{k=2}^L \left( \frac{k-1}{L} \right)^{\ell_k}~.
\end{equation}
In this limit, the speed of the shepherd reduces to $V_L=1/L$ and other
characteristics of the system similarly simplify.

A useful feature of our model is that it can be mapped onto the {\em zero
  range process} (ZRP) \citep{evans05} on a chain of $L$ sites with open
boundaries.  In this mapping, each knot corresponds to a site in the ZRP, and
the number $n_k$ of empty sites to the left of each knot corresponds to the
number of particles that occupy site $k$ in the ZRP.  The particles in the
ZRP multiply occupy each site, and only the top particle can hop at a rate
that is independent of the occupancy of the target site.  For example, knot
$k$ hopping to the right corresponds to a particle in the ZRP hopping from
site $k+1$ to $k$.  The top particle in this corresponding ZRP can hop: (i)
both to the left and to the right at rate 1 from each bulk site; (ii) out of
site $k=1$ at rate $\gamma$ and into $k=1$ at rate $\alpha$; (iii) either in
or out of site $k=L$ at rate 1.  The relevant feature of this mapping is that
the negative of the current entering the ZRP from the left exactly
corresponds to the speed of the shepherd.

Our shepherd model is a special case of the general ZRP that was solved by
\citet{levine05} and Eqs.~\eqref{prodanz}, \eqref{solz}, and \eqref{speed}
are special cases of their solution.  In addition to its average speed, we
are also interested in the diffusion coefficient of the shepherd. Due to the
equivalence of the models, the negative time integrated current $-J(t)$ of the ZRP corresponds to the position $x(t)=-J(t)$ of the shepherd in our model. We can use results of \citet{harris05} for the
fluctuations of the time-integrated current. From this work, the
large time asymptotic of the Laplace transform of the position of the shepherd is given
explicitly by (the large-$L$ limit of the expression below was obtained
previously in \citet{bodineau04,wijland05}):
\begin{equation}
\label{lap}
 \hat x(\lambda) \equiv \lim_{t\to\infty} \frac{1}{t} \log \langle e^{-\lambda x(t)} \rangle
 = \frac{(1-e^\lambda)(\alpha e^{-\lambda}-\gamma)}{1+\gamma L}~.
\end{equation}
From this equation, the speed of the shepherd \eqref{speed} can be obtained as $V_L=\partial_\lambda \hat x|_{\lambda=0}$, while the diffusion coefficient is 
\begin{equation}
  D_L =  \frac{1}{2}\, \left.
    \frac{\partial^2 \hat x}{\partial\lambda^2} \right|_{\lambda=0}=
  \frac{\alpha+\gamma}{2(1+\gamma L)}
\end{equation}
Again, the result is particularly neat, $D_L=(2L)^{-1}$, in the limit of
an infinitely biased shepherd.


\subsection{Ejection Time}

\begin{figure}[ht]
\centering
\includegraphics[width=0.45\textwidth]{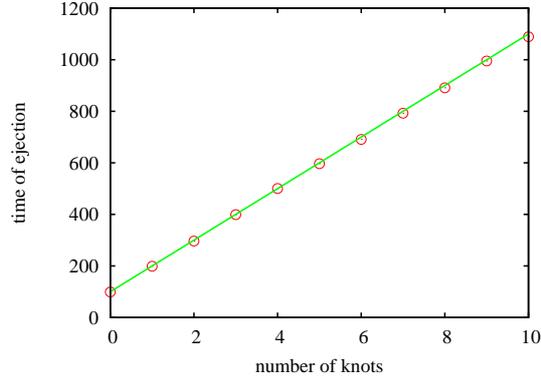}
\caption{Average ejection time for a polymer of length 100 as a function of
  the number of equidistant knots $L$.  The hopping rates of the shepherd are
  $\alpha=1$, $\gamma=2$.  The line is the function $100*(1+L)$ as given by
  Eq.~\eqref{time}. }
\label{fig:avtime}
\end{figure}

What is the average ejection time of a polymer with a finite density of
knots?  For concreteness, consider a polymer of length $N$, with a shepherd
at site 0, and $L$ equidistant knots initially at $x_i=iN/(L+1)$,
$i=1,\dots,L$.  If the knot density $\rho_\infty=L/N$ is sufficiently low
(roughly if $\rho_\infty\ll \gamma-\alpha$), the biased shepherd reaches each
knot close to the knot's initial position. The shepherd advances with the
localized flock of $k$ already-collected knots with asymptotic speed
$V_k=(\gamma-\alpha)/(\gamma k+1)$, see Eq.~\eqref{speed}, until the flock
reaches the $(k+1)^{\rm st}$ knot.  Consequently, the time for the shepherd
to reach the end of the polymer is
\begin{equation}
\label{time}
 t = \frac{N}{L+1}\sum_{k=0}^L \frac{1}{V_k} 
= \frac{N}{2}\, \frac{\gamma L+2}{\gamma-\alpha}
\end{equation}
Simulation results for this time are plotted in Fig.~\ref{fig:avtime} for a
polymer of length $N=100$ as a function of the number of knots when the bias
of the shepherd is weak ($\alpha=1, \gamma=2$).  Our result \eqref{time}
gives an excellent fit to the data even for this short polymer.

For a finite density of knots $\rho_\infty=L/N$, we also obtain the
asymptotic position of the shepherd from \eqref{time} as
\begin{equation}
\label{pos}
 x \simeq \sqrt{\frac{2(\gamma-\alpha)}{\gamma\rho_\infty}\,
   t\,\,}\longrightarrow \sqrt{2t/\rho_\infty}\qquad {\alpha\to 0}~.
\end{equation}
Notice the lack of dependence on the hopping rate $\gamma$ of the shepherd in
the asymmetric limit of $\alpha\to 0$.

When a real virus is ejected from the capsid, the pressure should decrease during this process \citep{GK09}. This effect can be modeled as a decreasing bias $\gamma$ for the shepherd. Since the bias is so enormous in the capsid  \citep{GK09}, even the decreased bias can be considered as large bias in our model. This argument is supported by the fairly uniform rate of virus ejection in simulations of \citet{matthews09}. The ejection time \eqref{time} for large bias becomes $t=NL/2$, that is independent of the rates $\alpha$ and $\gamma$. Knowing this ejection time experimentally would give us an estimate for the hopping rates of the knots, (which we used to rescale time with). Unfortunately, we do not know of such experiments.

\section{Finite Density of Knots} 
\label{cont}

When the number of knots is large, it is more convenient to study the knot
density profile by a continuum theory.  While a finite knot density is
perhaps not of direct relevance to the polymer ejection problem, this
limiting situation leads to an appealing exclusion process that can be solved
in a simple way by applying a scaling approach to account for the effective
moving boundary condition caused by the motion of the shepherd \citep{Crank}.

\begin{figure}[ht]
\centering
\includegraphics[width=0.45\textwidth]{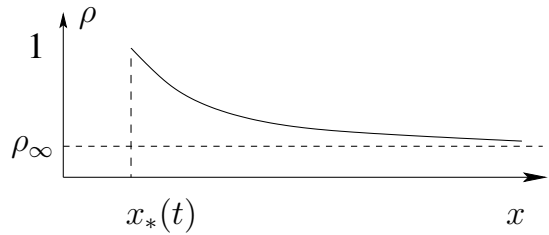}
\caption{ Sketch of the knot density as a function of position along the
  chain in the continuum limit. }
\label{continuum}
\end{figure}

Let $\rho(x,t)$ denote the continuum knot density at position $x$ and time
$t$, and let $x_*(t)$ denote the position of the shepherd
(Fig.~\ref{continuum}).  Initially the shepherd is at $x_*=0$ and it hops to
the right with rate 1 whenever its right neighbor is vacant.  In the
continuum limit, the knot density $\rho(x,t)$ satisfies the diffusion
equation
\begin{equation}
\label{DE}
\frac{\partial \rho}{\partial t} = \frac{\partial^2 \rho}{\partial x^2}~.
\end{equation}
The diffusion coefficient $D=1$ since the hopping rate of each knot equals 1
and the exclusion plays no role in the overall density when the hopping is
symmetric \citep{schutz00}.

As the shepherd advances, knots accumulate in front of it and the knot
density increases monotonically from a value $\rho_\infty$ as $x\to\infty$ to
the value 1 at $x_*$.  The shepherd advances only when the adjacent site is
empty.  In the continuum limit, the analog of this advancement rule is
\begin{equation}
\label{Stefan}
\frac{dx_*}{dt} = - \frac{\partial \rho}{\partial x}\Big|_{x=x_*}.
\end{equation}
That is, the probability that there is a gap in front of the shepherd equals
the local knot density gradient.  Thus we must solve the diffusion equation
\eqref{DE} in the region $x_*(t)<x<\infty$ subject to the initial condition
$\rho(x>0,0) = \rho_\infty$ and the boundary condition $\rho(x_*,t>0) = 1$.

This type of moving boundary-value, or Stefan, problem \citep{Crank,L87} can
be solved by a scaling approach.  Let us assume that the density profile
approaches the scaling form
\begin{equation}
\label{scaling}
\rho(x,t) = f(\xi), \quad \mathrm{where} \qquad \xi=\frac{x}{x_*}-1\,.
\end{equation}
Using this scaling form, the derivatives of the density are:
\begin{equation*}
\frac{\partial \rho}{\partial t} =\! -\frac{\dot x_*}{x_*}\,(1\!+\!\xi)f',
\quad \frac{\partial \rho}{\partial x} = \frac{1}{x_*}\,f', 
\quad \frac{\partial^2 \rho}{\partial x^2} = \frac{1}{x_*^2}\,f''\,.
\end{equation*}
Using these, the diffusion equation \eqref{DE} can be written in the
separated form
\begin{equation*}
\dot x_*x_* = - \frac{f''(\xi)}{\,(1\!+\!\xi)f'(\xi)},
\end{equation*}
where $A$ is a separation constant.  Equation~\eqref{Stefan} for the motion
of the boundary gives
\begin{equation}
\label{Stefan*}
x_*\,\dot x_* = -f'(0)\,,
\end{equation}
which both fixes the separation constant $A=-f'(0)$ and implies that the
boundary point advances in time as
\begin{equation}
x_* = \sqrt{2A t}\,.
\end{equation}

In scaled coordinates the differential equation for the density profile now
becomes $f''(\xi)=-A(1+\xi)f'(\xi)$.  Integrating this equation, subject to
the boundary condition $\rho(x_*,t>0) = 1$ (or $f(0)=1$), yields
\begin{equation}
f(\xi)=1-A\int_0^\xi d\eta\, e^{-A(\eta+\eta^2/2)}~.
\end{equation}
To eliminate the unknown constant $A=-f'(0)$, we use the fact that
$f(\infty)=\rho_\infty$ to obtain the relation
\begin{eqnarray*}
  \rho_\infty &=& 1 -A\int_0^\infty d\eta\,\,e^{-A(\eta+\eta^2/2)}\\
&=& 1 -\sqrt{\pi A/2}\,\, e^{A/2}\,\,{\rm erfc}(\sqrt{A/2}).
\end{eqnarray*}
that (implicitly) determines $A$ as a function of $\rho_\infty$; here
$\mathrm{erfc}$ is the complementary error function \citep{AS}.  Using the
asymptotic forms of the error function, we can extract the limiting behaviors
of the separation constant $A$, from which
the position of shepherd is given by
\begin{equation}
x_* \simeq  \sqrt{t}\times
\begin{cases}
\sqrt{2/\rho_\infty}             & \rho_\infty\downarrow 0\\
2\pi^{-1/2}(1-\rho_\infty)   & \rho_\infty\uparrow 1~.
\end{cases}
\end{equation}
Notice that the limiting expression for $\rho_\infty\to 0$ reproduces
Eq.~\eqref{pos} that we obtained from the exact discrete solution.

\section{Summary}

We modeled the ejection of a knotted flexible polymer from a capsid in terms
of a hybrid dynamical model that consists of a single asymmetric exclusion
process (representing the boundary point on the polymer between the interior
and exterior of the capsid) that interacts with a gas of symmetric exclusion
processes (the knots).  In the absence of any external forces, the knots
reptate symmetrically along the chain.  The effect of the osmotic pressure is
modeled as the single asymmetric exclusion process that shepherds the knots
to the opposite end of the chain where their entanglement is released.  Once
all the knots have unraveled, the polymer can be ejected.

The ejection speed $V_L$ can be solved exactly for the $L+1$-particle system
that consists of the shepherd and $L$ knots and the result is remarkably
simple for a strongly-biased shepherd: $V_L=1/L$.  The underlying exclusion
process can also be solved for a finite density of knots by a continuum
description.  Here the amount of the chain that is ejected at time $t$ is not
proportional to $t$, but rather to $\sqrt{t}$.  This behavior matches that of
the discrete solution in the limit of a strongly-biased shepherd.  In all of
our theoretical modeling, we use constant bias throughout the entire ejection
process.  We argue that the motion remains in the strong-bias regime
throughout the ejection process due to the extremely large initial pressure
in the capsid.  As long as the system remains in this strong-bias limit, the
motion of the shepherd and the knots are not influenced by the decrease in
the ejection force and it is appropriate to treat the bias as constant.

One final point is that we have assumed that all knots are identical.  It may
be worthwhile to extend our model to the physical more realistic case where
knots are non-identical, so that their diffusivities are random variables.

\section*{Acknowledgments}

We thank Gleb Oshanin for useful correspondence about lattice gases.  
We are grateful for financial support from NIH grant R01GM078986 (TA), 
NSF grants CCF-0829541 (PLK) and DMR0535503 (SR).

\end{document}